# Pre-stressed Sub-surface Contribution on Bulk Diffusion in Metallic Solids

Laura Raceanu[1,a], Virgil Optasanu[1,b], Tony Montesin[1,c] and Nicolas Creton[1,d]

[1]ICB, UMR 5209 CNRS, 9 Avenue Alain Savary, BP 47870, 21078 DIJON

[a]laura.raceanu@u-bourgogne.fr, [b]virgil.optasanu@u-bourgogne.fr, [c]tony.montesin@u-bourgogne.fr, [d]nicolas.creton@u-bourgogne.fr

**Abstract.** Our recent modelling works and corresponding numerical simulations realized to describe the $UO_2$ oxidation processes confirm the theory showing that an applied mechanical strain can strongly affect the local oxygen diffusion in a stressed solid. This result allows us to assume that stress field, previously applied at the surface of a metallic sample on several microns, will delay the degradation during its oxidation. Considering this hypothesis, we implemented a FEM simulation code developed in our laboratory to numerically investigate some different stress fields applied on a sample sub-surface, that might significantly modify the volume diffusion of oxygen during the oxidation process. The results of our simulations are presented and discussed from the perspective to study the consequences of a surface mechanical treatment on the durability of a metallic material.

**Introduction**

Last decade, many works concerning interactions between mechanical stresses and chemical diffusion have been done in order to understand the mechanisms and their consequences on the ageing of metallic or ceramic solids [1] [2] [3].

Thus, under particular pressure and temperature metals can react in contact with the environment and lead to corrosion. One of the most important mechanisms of these behavior is the species diffusion which gives composition gradients in the solid producing strong gradients of mechanical stresses with important feedback on the diffusion itself [4].

At the same time, under mechanical loads, metallurgical heterogeneity may produce species diffusion and eventually form segregation zones which can be at the origin of local embrittlement [5]. Many works point the active role of stresses on the material ageing especially on their negative consequences leading to the damaging of structures. On the opposite side, the present work consist in use the strong coupling between mechanical stresses and chemical diffusion in order to increase the life expectancy of structures by manipulating the internal stresses inside the solid.

Our works are based on a both theoretical and experimental study about $UO_2$ oxidation. The conclusions of this study highlight the role of the coupling effect of stresses and oxygen bulk diffusion on the damage of the $U_3O_7$ oxide layer. The numerical tools developed in our laboratory were now used to identify the various theoretical stress profiles which can influence the diffusion of the oxygen into Zr.

The present work is the first step of an identification study for defining an experimental protocol for a surface treatment designed to impose a given stress field under the material surface. This stress could help to improve the corrosion resistance and subsequently the life expectancy of a metallic structure.

**Stress-diffusion coupling in $UO_2/U_3O_7$ system**

Recently, the behaviour of the couple $UO_2/U_3O_7$ ($UO_2$: uranium dioxide for nuclear industry), submitted to high temperature oxidation (> 350°C), was studied in order to identify the critical $U_3O_7$ oxide layer thickness, beyond which a first macroscopic crack appears. The knowledge of this yield is essential to better control fuel fabrication and safety of spent fuel dry storage in nuclear industry [6].

The experiments realised by Rousseau et al. [7] on oxidized $UO_2$ single crystals, underlined an increase of the average stress inside the oxide layer during the growth of this one. Under certain conditions of temperature (> 350°), this increase is accompanied by the formation of $U_3O_7$ domains, organised according to preferential crystallographic directions at the $UO_2$ interface. This specific organisation would be at the origin of the oxide layer cracking, when its growth reaches a given yield.

Some numerical simulations, respecting the experimental conditions implemented by Rousseau revealed, inside the substrate but close the interface, the presence of both a stress gradient and a composition gradient, at once in limited depth and in a parallel plane to the substrate/oxide interface (cf. figure 1).

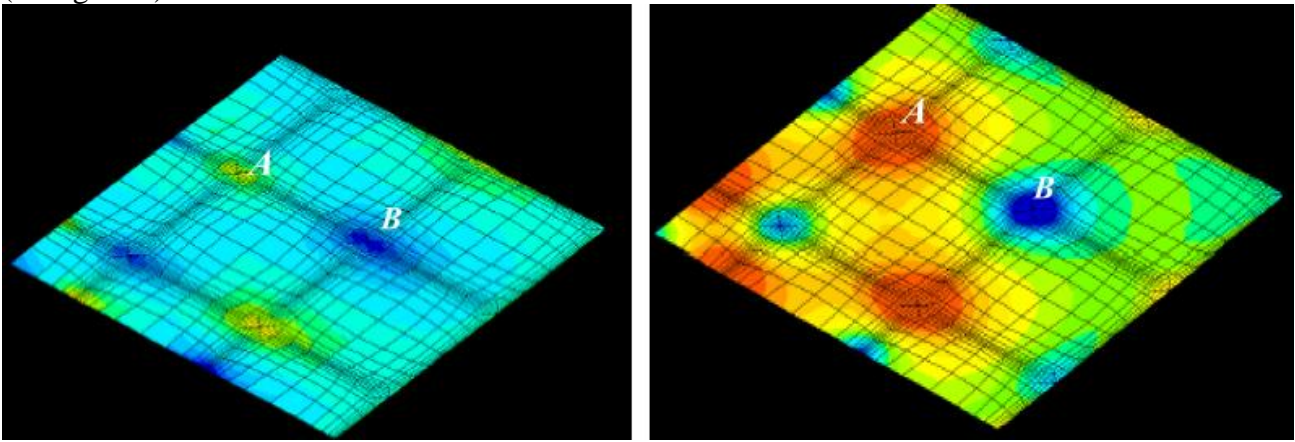

Figure 1. sectional view of a $UO_2$ single crystal's extreme surface, in a plane parallel to the $UO_2/U_3O_7$ interface (x and y axis). Stresses (a) and oxygen composition (b) are due to the formation of several $U_3O_7$ crystals, oriented under particular directions.

The composition field, highly depending on the stress field, brings up some matter volumes enriched in oxygen whereas some others are impoverished. As a result, a heterogeneous growth of the oxide into the interface plane appears which leads to a heterogeneous stress repartition into the oxide, inducing its crack when the layer reaches a limit thickness.

The finite element simulation widely developed by Creton et al. [3] [8] relies on a macroscopic and thermodynamic formulation of the matter transport equation, including mechanical and epitaxial strains. From a mechanical point of view, the solid behaviour is supposed to be purely elastic and anisotropic. In the volume, the matter flux relative to the diffusing species (oxygen in our case), is given by a generalized Fick's law in which the diffusion coefficient depends on the stresses:

$$\vec{J}_O = -D_O \left[ \vec{\nabla} c_O + \frac{N c_O V_O(\sigma)}{RT} \vec{\nabla} \sigma \right], \quad (1)$$

where $D_0$ the diffusion coefficient in stress-free state, $M_0$ represents the molar mass of the $UO_2$, $\eta_{ij}$ is the chemical expansion coefficient, $c$ is the oxygen concentration which is dissolved in $UO_2$ lattice, $\sigma$ is the mechanical stress, $T$ is the temperature and $R$ is the universal constant of gases.

The stress profiles observed in $UO_2$ substrate, close to substrate/oxide interface, and at two singular points A and B (see figure 1b) are presented on figure 2. The z-axis materializes the thickness of the substrate, and the origin of the basis is set at the substrate/oxide internal interface.

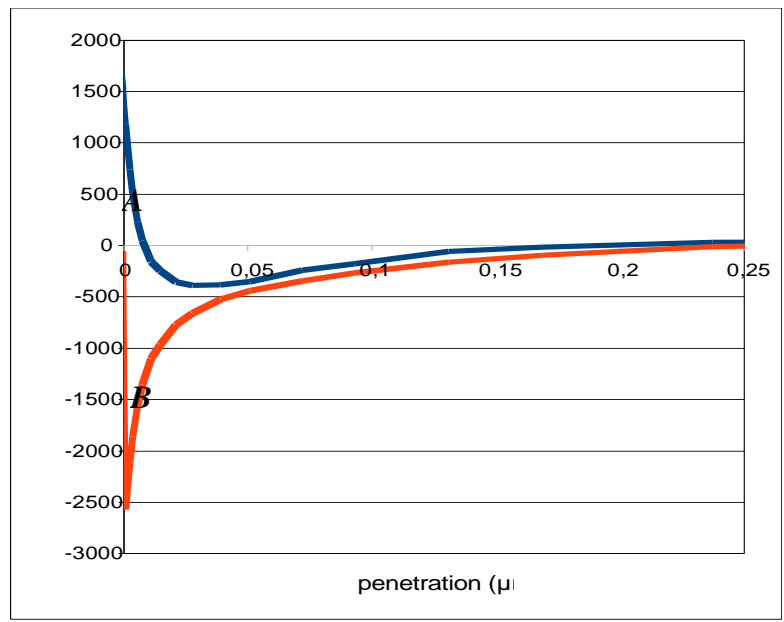

Figure 2. $\sigma_{xx}$-stress profiles into $UO_2$ at the substrate/oxide internal interface presented figure 1 on two singular points A and B.

Each one of these two points are representative of a domain of high (point A) or low (point B) concentration in oxygen, under stress influence. The strong gradients observed on figure 2 take their origin from:
- the oxygen dissolution into $UO_2$ substrate,
- the strains due to the accommodation of $UO_2$ and the $U_3O_7$ lattice crystals,
- the heterogeneity of $U_3O_7$ crystalline domains.

This modelling confirms the conclusions resulting from Rousseau's experimental studies. Furthermore, it reveals that by imposing a strain field at the surface of a solid subject to both oxidant environment and bulk diffusion of species, it's possible to act on the concentration evolution of this one, and so to modify the growth rate of the oxide layer formed.

### Numerical approach of an "active" stress profile

Could a pre-stress (applied to the extreme surface of a sample that will be in a second time subject to a corrosive environment) be able to affect the chemical process?

In order to propose an answer to this question, several numerical simulations were explored. A parallelepiped sample, 15x15x2 mm³, of pure Zr was first subject to a pre-stress in extreme surface of the solid (few microns in depth). The model is based on a stress/diffusion coupling interactions according to the approach developed in our laboratory (see equation 1). In first approximation, the material is supposed to be elastic and isotropic. The sample is heated up to 700° C, which is in the range of temperature corresponding to an oxygen insertion in the crystal lattice during the diffusion.

The choice of the Zr is mainly due to the abundance of the physico-chemical and mechanical data available in the literature. This material was also widely studied in the past in our laboratory [4][9][10].

Two profiles of pre-stress, covering a depth of 5μm were considered: a tensile one (figure 3, curve ❶) or a compressive one (having the same absolute value as the curve ❶). Each one was applied parallel to the oxygen adsorbent surface. The maximum level of this pre-stress corresponds to the Zr yield strength (180 MPa). In order to develop our analysis, two other initial conditions, were also considered: short range oxygen diffusion (figure 3 curve ❷) or long range oxygen diffusion (figure 3 curve ❸) into the solid. These special conditions are applied before the pre-stress.

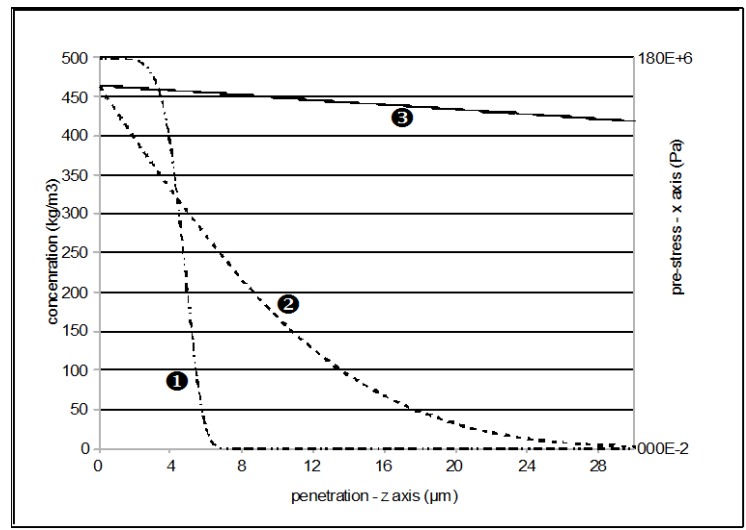

Figure 3. Three initial conditions of composition an stress ($\sigma_{xx}$) imposed to the solid: - oxygen saturation : 464 kg/m$^3$ - Maximum stress level : 180 MPa (Zr yield strength).

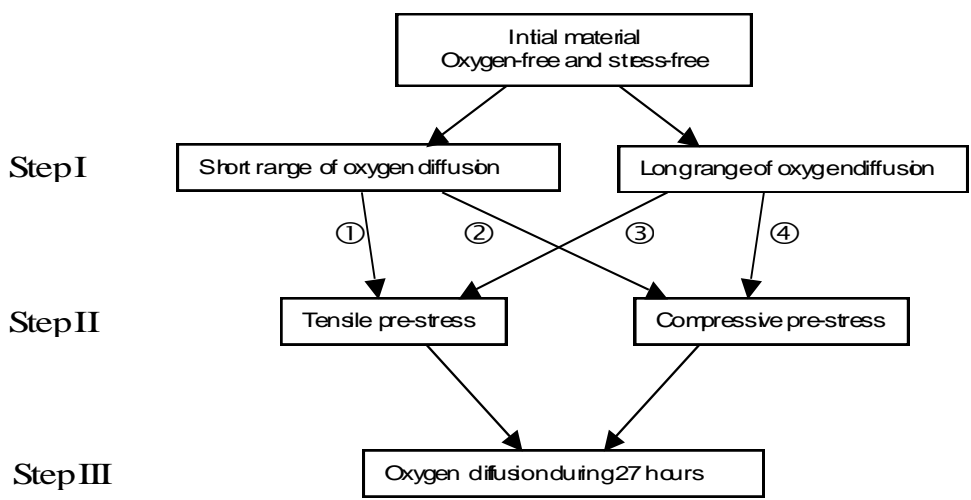

Figure 4. Numerical model scheme

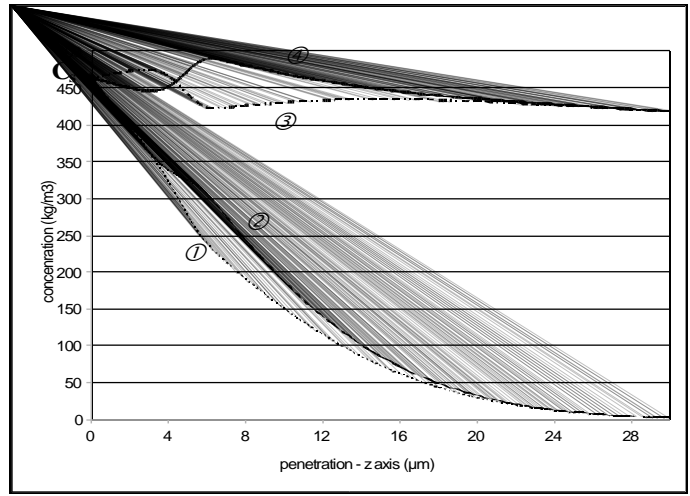

Figure 5. Computed composition profiles in Zr according to different initial states of the material:
① Short range diffusion and tensile pre-stress,
② Short range diffusion and compressive pre-stress
③ Long range diffusion and tensile pre-stress
④ Long range diffusion and compressive pre-stress

The protocol of our numerical implementation is described by the following scheme (figure 4). Steps I and II configure the initial conditions used for the final calculation (step III), which corresponds to dissolution of oxygen into Zr during 27 hours. Figure 5 presents the computation results.

It appears, in the case ①, that the pre-stress affects the oxygen profile in the metal, alongside the stress front (position z = 5μm). As the curves ① and ② are superposed close to $C_s$ saturation point (464 kg/m3), the stress consequences (compression or tension) on oxygen fluxes between z=0 and z=5μm should not be perceptible after 27 hours of oxygen dissolution. In the case of long range oxygen diffusion, the pre-stress influences the distribution of the composition not only in the zone where the material was subject to a mechanical strain (z=0 and z = 5μm) but also beyond. We remark here an accumulation of oxygen at the right side of stress front (at z = 5μm) in the case of a compressive stress. In the case of tensile stress, oxygen would accumulate at the initial mechanically strained zone. If an oxide grows, the internal interface displacement will not be done with the same speed, if the material was previously pre-stressed under a tensile stress (acceleration) or a compressive stress (slowdown).

The figure 5 clearly shows the competition existing between the stress gradient and the composition gradient in the oxygen diffusion process. So, the smaller is the composition gradient, the more rapidly perceptible is the stress effect.

**Summary and prospects**

The numerical implementation of a stress/diffusion coupling in uranium dioxide during its oxidation, allowed us to underline the influence of the stress on the oxide $U_3O_7$ layer growth. The stress fields susceptible to act on the oxygen dissolution in the material present the special characteristic of a strong variation in a very low depth of the substrate.

Furthermore, numerical simulations of oxygen dissolution into Zr were accomplished by imposing, as initial conditions, different stress profiles close to those obtained at the beginning of this work (case of $UO_2$). The composition fields resulting from the simulation (during 27 hours) of oxygen diffusion in pure Zr make obvious a non-negligible impact of the pre-stresses profiles introduced. These results allow us to believe in a possible control of the chemical process (dissolution- oxidation) in the case of a system subject to bulk diffusion fluxes. To validate these conclusions, it is necessary to confront them to different experimental approaches.

Several technologies can be used to impose a pre-stress to the surface of a strained solid, according to profiles we numerically explored. They are related to thermo-mechanical surface treatments (laser shock and mechanical attrition).

Three particular pre-stress profiles will be soon the object of numerical and experimental investigations. They correspond to residual stresses relating to mechanical (SMAT [11] and shot-peening [12]) or chemical (implantation of noble gas) surface treatments technologies. The applications will concern different materials/diffusing species coupling: Zr – O, Steel – H and $UO_2$ – O. Furthermore, aware that a mechanical treatment drives to plastic strains, it will be necessary to evaluate the consequences of the local modification on the diffusion paths of the concerned species.

**References**

[1] J. Godlewski, P. Bouvier, G. Lucazeau, and L. Fayette, ASTM (2000), p. 877-900
[2] A. Adrover, M. Giona, L. Capobianco, P. Tripodi, and V. Violante: J. Alloys Comp. Vol. 358 (2003), p. 268.
[3] N. Creton, V. Optasanu, T. Montesin, S. Garruchet, and L. Desgranges : Def. Diff. Forum Vol. 289-292 (2009), p. 447.
[4] J. Favergeon, T. Montesin, and G. Bertrand: Oxidat. Met. Vol. 64 (2005), p. 253.
[5] AK. Jha, S. Manwatkar, and K. Sreekumar: Eng. Failure Anal. Vol. 17-4 (2010), p. 777.
[6] RJ. McEachern and P. Taylor: J. Nucl. Mater. Vol 254 (1998), p. 87.
[7] G. Rousseau, L. Desgranges, J.C. Niepce, G. Baldinozzi, and J.F. Berar: J. Phys. IV, Vol. 118 (2004), p. 127.
[8] N. Creton, V. Optasanu, S. Garruchet, L. Raceanu, T. Montesin, L. Desgranges, and S. Dejardin: Def. Diff. Forum Vol. 297-301 (2010), p. 519.
[9] C. Valot, D. Ciosmak, and M. Lallemant: Solid State Ionics Vol. 101 (1997), p. 769.
[10] S. Garruchet, T. Montesin, H. Sabar, M. Salazar, and G. Bertrand : Mater. Sci. Forum Vol. 461-464 (2004), p. 611.
[11] L. Zhang, Y. Han, and J. Lu: Nanotechnol. Vol. 19 (2008), p. 165706.
[12] T. A. Hayes, M. E. Kassner, D. Amick, and R. S. Rosen: J. Nucl. Mater. Vol. 246 (1997), p. 60.